\documentclass[letterpaper, 10 pt, journal]{ieeeconf}

\IEEEoverridecommandlockouts                              
                                                          
\usepackage{amsmath}
\usepackage{amssymb}
\usepackage{tabularx}
\usepackage{xfrac}
\usepackage[final]{pdfpages}
\usepackage{algorithm}
\usepackage[noend]{algpseudocode}
\usepackage{booktabs}
\usepackage[short]{optidef}
\usepackage{soul,xcolor}
\usepackage{import}
\usepackage{bm}
\usepackage{tablefootnote}
\usepackage{physics}

\usepackage{units}
\usepackage{arydshln}
\usepackage{lipsum}

\usepackage{url}
\usepackage[hidelinks]{hyperref}
\usepackage{balance}

\usepackage{graphicx}
\usepackage{comment} 

\usepackage{cite}
\usepackage{mathtools,cuted}

\usepackage[capitalise]{cleveref}

\usepackage{caption}
\usepackage{subcaption}

\usepackage{nomencl}
\makenomenclature

\Crefname{equation}{Eq.}{Eqs.}
\Crefname{figure}{Fig.}{Figs.}
\Crefname{tabular}{Tab.}{Tabs.}
\Crefname{line}{L.}{L.}

\renewcommand{\nompreamble}{The next list describes several symbols that we used throughout the paper and their explanation.}

\newlength\figureheight
\newlength\figurewidth

\newcommand{\Pkp}{P_{\text{kp}}} 
\newcommand{\Vkp}{V_{\text{kp}}} 
\newcommand{\Vki}{V_{\text{ki}}} 
\newcommand{\Cp}{C_{\text{p}}} 
\newcommand{\Cv}{C_{\text{v}}} 
\newcommand{\Vff}{V_{\text{ff}}} 
\newcommand{\Aff}{A_{\text{ff}}} 

\newcommand{\data}{\mathcal{D}}

\newcommand{\mypar}[1]{\textbf{#1.}}
\newcommand{\safeopt}{\textsc{SafeOPT}}
\newcommand{\stageopt}{\textsc{StageOPT}}
\newcommand{\goose}{\textsc{GoOSE}}

\newboolean{include-notes}
\setboolean{include-notes}{true}

\newcommand{\matteo}[1]{\ifthenelse{\boolean{include-notes}}{{\color{orange} \textbf{Matteo}: #1}}{}}
\newcommand{\christopher}[1]{\ifthenelse{\boolean{include-notes}}{{\color{green} \textbf{Christopher}: #1}}{}}
\newcommand{\alisa}[1]{\ifthenelse{\boolean{include-notes}}{{\color{red} \textbf{Alisa}: #1}}{}}
\newcommand{\andreas}[1]{\ifthenelse{\boolean{include-notes}}{{\color{blue} \textbf{Andreas}: #1}}{}}

\newcommand{\R}{\mathbb{R}}

\newcommand{\X}{\mathcal{X}}

\newcommand{\Si}{\mathbf{\Sigma}}

\newcommand{\lcb}{\mathrm{lcb}}
\newcommand{\ucb}{\mathrm{ucb}}
\newcommand{\acq}{\mathrm{acq_{t}}}

\newcommand{\argmin}{\operatornamewithlimits{argmin}}

\begin{document}


\title{ \LARGE \bf Safe Risk-averse Bayesian Optimization for Controller Tuning}
\author{Christopher K{ö}nig$^{1}$, Miks Ozols$^{2}$, Anastasia Makarova$^{3}$, \\ 
Efe C. Balta$^{2}$, Andreas Krause$^{3}$, Alisa Rupenyan$^{1,2}$ 
\thanks{This project was funded by the Swiss Innovation Agency, grant Nr. 46716, and by the Swiss National Science Foundation under NCCR Automation, grant Nr. 180545. \newline
$^{1}$ Inspire AG, Zurich, Switzerland \newline
$^{2}$ Automatic Control Laboratory, ETH Zurich, Switzerland \newline
$^{3}$ Learning \& Adaptive Systems group, ETH Zurich, Switzerland }}

\maketitle


\begin{abstract}
Controller tuning and parameter optimization are crucial in system design to improve both the controller and underlying system performance. Bayesian optimization has been established as an efficient model-free method for controller tuning and adaptation. Standard methods, however, are not enough for high-precision systems to be robust with respect to \emph{unknown} input-dependent noise and stable under safety constraints.   In this work, we present a novel data-driven approach, RaGooSe, for safe controller tuning in the presence of heteroscedastic noise, combining safe learning with risk-averse Bayesian optimization. We demonstrate the method for synthetic benchmark and compare its performance to established BO-based tuning methods. We further evaluate RaGoose performance on a real precision-motion system utilized in semiconductor industry applications and compare it to the built-in auto-tuning routine.
\end{abstract}
\section{Introduction}
High-precision, high-stakes systems operation often requires risk-averse parameter selection, even if this results in a trade-off with the best expected performance of the systems. The standard approach in tuning the control parameters of such systems assumes constant observation noise, and tuning is often based on a model-based or rule-based procedure, thus limiting the system's performance. The goal of this work is to demonstrate a novel \emph{flexible, risk-averse} approach to find optimal control parameters under input-dependent noise, based only on observation of the system performance.

Optimizing performance indicators derived from system data for tuning of control parameters has been explored through various approaches, e.g., iterative feedforward tuning \cite{IterTuning_Li2018}, variable gain selection \cite{VarGain_Li2015}, data-driven feedforward learning \cite{FeedforLearning_Song2020}. The performance criteria can be represented by features in the data measured during system operation at different values of the controller parameters \cite{khosravi2020cascade}. Then, the low-level controller parameters can be optimized to fulfill the desired performance criteria. 
Bayesian optimization (BO) \cite{Mockus} denotes a class of sample-efficient, black-box optimization algorithms that have been used to address a wide range of problems, see \cite{shahriari2015taking} for a review. BO has been successfully demonstrated in controller tuning with custom unknown objectives and constraints in the existing literature~\cite{Duivenvoorden,Marco2016}.

An important aspect for practical applicability of BO is safety.
In this work, we follow the definition for \emph{safety} from \cite{berkenkamp2021bayesian} as optimization under constraints, separate from the objective. 
BO has been combined with heteroscedastic noise for controller optimization \cite{guzman2020heteroscedastic} to find optimal hyperparameters of a stochastic model predictive controller, however, the safety aspect of the approach was  not studied. Our approach considers safety of the underlying system and combines aspects of safe BO and risk-averse BO. This ensures that there are no constraint violations during the optimization, and enables the application of the approach to continuous optimization, e.g., \cite{koenig2021, brunzema2022controller} concerning controller parameter adaptation for systems operating under changing conditions. 

\mypar{Contribution} 
 In this work, we make the following contributions: (1) we propose a Bayesian optimization algorithm for safe policy search in the presence of heteroscedastic noise. The noise affects the surrogate models used in the data-driven constrained optimization procedure, which are built using features from the underlying noisy data, and the exploration-exploitation strategy. (2) We demonstrate the approach for a benchmark problem with safety constraints and compare the achieved performance to that achieved without accounting for the noise based on \cite{koenig2021}, and that achieved with another constrained BO approach \cite{Gardner}. (3) We demonstrate the approach on a numerical simulation of a high-precision motion system, and apply it in practice to tune the controller on the real system (Fig. \ref{fig:argussetting}), used in semiconductor manufacturing, in comparison with the industry standard tuning approach, via the built-in auto-tuner in the system.
 \vspace{-5 mm}


\section{Related work}

\mypar{BO for controller tuning} 
BO has been applied in the tuning of various  types of controllers for fast motion systems. BO-based tuning for enhanced performance has been demonstrated for cascade controllers of linear axis drives, where data-driven performance metrics have been used to intentionally increase the traversal time and the tracking accuracy while reducing vibrations \cite{Khosravi2020}. 
In model predictive control (MPC), instead of adapting the controller for the worst-case scenarios, the prediction model can be selected to provide the best closed-loop performance by tuning the parameters in the MPC optimization objective for maximum performance \cite{Piga_2018, sorourifar2020data, rupenyan2021performance}.

\mypar{Safe BO} To ensure optimal control performance for time-varying systems, controllers should adapt their parameters over time, while avoiding unsafe, or unstable parameters. Especially in high-precision motion systems, even one iteration with excessive vibrations is not allowed during operation. Thus, learning the parameters of cascaded controllers with safety and stability constraints is needed to ensure that only safe parameters are explored~\cite{koenig2021, brunzema2022controller}. The SafeOpt algorithm \cite{sui} achieves safety, but is inefficient due to its exploration strategy \cite{berkenkamp2021bayesian}.  
This has been addressed in  \cite{fiducioso2019safe}, where  the safe set is not actively expanded, which may compromise optimality but works well for the considered application. Another approach applied to controller tuning, is to add a safety-related log-barrier in the cost, as demonstrated in \cite{khosravi2022safety}, to reduce constraint violations as compared to an implementation based on BO with inequality constraints \cite{Gardner}.
An efficient way to enable safe exploration is proposed by GoOSE \cite{turchetta2019safe}, ensuring that every input to the system satisfies an unknown, but observable constraint. For controller tuning, GoOSE unifies time-varying Gaussian process bandit optimization \cite{bogunovic2016time} with multi-task Gaussian processes \cite{swersky2013multi}, and with efficient safe set search based on particle swarm optimization, enabling its application in the adaptive tuning of motion systems  \cite{koenig2021}.

 \mypar{Tuning in the presence of heteroscedastic noise} Input-dependent noise and lack of repeatability are common in robotic and motion systems. The main challenge is that both the objective and noise are unknown. Taking it into account in controller design leads to performance improvement, as demonstrated in the MPC-based controller design of robotic systems. For example, \cite{guzman2020heteroscedastic} and \cite{guzman2022bayesian} account for heteroscedastic noise via an additional flexible parametric noise model. These approaches, however, utilize a restrictive model of the noise variance. \looseness-1
 
 Alternatively, \cite{makarova2021risk} focus on the risk associated with querying noisy points and proposing RAHBO, a flexible approach enjoying theoretical convergence guarantees. In contrast to the standard BO paradigm that optimizes the expected value only and might fail in the risk-averse setting, RAHBO trades the mean and input-dependent variance of the objective, while learning both the objective and the noise distribution on the fly. To this end, RAHBO extends the concentration inequalities in the case of heteroscedastic noise \cite{kirschner2018information} to the setting of unknown noise following the optimism-under-uncertainty principle \cite{Srinivas_2012}. \looseness-1

\section{Preliminaries and Problem statement}\label{sec:problem_statement}
This section introduces the optimization problem of interest and related preliminaries.


\textbf{Classical optimization objective.} 
Consider a problem of sequentially optimizing a fixed and unknown objective $f: \X \rightarrow \R$ over the compact set of inputs $\X\subset\R^d$:
\begin{equation}
    \min_{x \in \X} f(x).
    \label{eq:bboopt}
\end{equation}
At every iteration $t$, for a chosen input $x_t \in \X$, one observes a noise-perturbed evaluation $y_t$ resulting into the dataset $\textstyle{\data_t=\{(x_i,y_i)\}_{i=1}^t}$, with: 
\begin{equation} \label{eq:observational_model}
    y_t = f(x_t) + \varepsilon(x_t),
\end{equation}
where $\varepsilon(x_t)$ is zero-mean noise, independent across different time steps $t$. 
In the standard setup, the noise is assumed to be identically distributed over the whole domain $\X$, i.e., $\varepsilon_t \sim \mathcal{N}(0,\rho^2)$. This, however, might be restrictive in practice, leading to sub-optimal solutions \cite{makarova2021risk}. Here we present the generalized version of the optimization with $\varepsilon_t \sim \mathcal{N}(0,\rho^2(x_{t}))$ and later describe why it is important. 

Bayesian optimization employs two main ingredients: \emph{probabilistic model} (for uncertainty quantification) and \emph{acquisition function} (for sampling policy of the next inputs).




\textbf{Probabilistic model.} 
Gaussian processes (GPs) \cite{Rasmussen} provide a distribution over the space of functions commonly used in non-parametric Bayesian regression. Posterior GP mean and variance denoted by $\mu_t(\cdot)$ and $\sigma^2_t(\cdot)$, respectively, are computed based on the previous measurements $y_{1:t} = [y_1, \dots, y_t]^\top$ and a given kernel $\kappa(\cdot,\cdot)$ :
\begin{align}
    \mu_t(x) &= \kappa_t(x)^T(K_t + \Si_t)^{-1} y_{1:t} \label{eq:GPmean},\\
     \sigma^2_t(x) &=  \big(\kappa(x,x) - \kappa_t(x)^\top(K_t + \Si_t)^{-1}\kappa_t(x) \big) \label{eq:GPvar},
\end{align} 
where $\Si_t = \text{diag}(\rho^2(x_1), \dots, \rho^2(x_t))$, $(K_t)_{i,j} = \kappa(x_i, x_j)$, $\ \kappa_t(\cdot)^T = [\kappa(x_1, \cdot), .., \kappa(x_t, \cdot) ]^T$, $f \sim GP(\mu_{0}, \lambda^{-1}\kappa)$ is a prior.

As long as the GP-based model of the objective $f$ is well-calibrated, we can use it to construct high-probability confidence bounds for $f$ \cite{Srinivas_2012}. In particular, as long as $f$ has a bounded norm in the reproducing kernel Hilbert space (RKHS) associated with the covariance function $\kappa$ used in the GP, $f$ is bounded by lower and upper confidence bounds  $\lcb^{f}_{t}(x)$ and $\ucb^{f}_{t}(x)$, respectively,
\begin{equation}
    \begin{aligned}
    \lcb^{f}_{t}(x) &= \mu_{t}(x) - \beta^{f}_{t} \sigma^{f}_{t}(x), \\
    \ucb^{f}_{t}(x) &= \mu_{t}(x) + \beta^{f}_{t} \sigma^{f}_{t}(x),\\
    \end{aligned}
    \label{eq:lcb_ucb}
\end{equation}
where $\beta^{f}_{t} > 0$ is a parameter that ensures validity of the confidence bounds \cite{kirschner2018information}.

\textbf{Acquisition function} $\mathrm{acq_t}: \X \rightarrow \R$
 expresses the informativeness of an input $x$ about the location of objective optimum, given the probabilistic model of $f$. The goal of the acquisition function is
to trade off \emph{exploration}, i.e., learning about uncertain inputs, and \emph{exploitation}, i.e., choosing inputs that lead to low cost function values. Thus, BO
reduces the original black-box optimization problem to cheaper problems $x_t = \argmin_{x \in \X}\mathrm{acq_t}(x)$.  One of the most popular acquisition functions, GP-LCB \cite{Srinivas_2012}, uses the so-called principle of \emph{optimism in the face of uncertainty}. The idea is to rely on an optimistic guess of $f(x)$ via the lower confidence bound in \cref{eq:lcb_ucb} and choose $x_t$ with the lowest guess. \cref{alg:training} provides the standard BO loop. \looseness-1
\begin{algorithm}[ht]
\caption{Bayesian Optimization loop}
 \label{alg:training}
    \begin{algorithmic}[1] 
       
        \State Prior $f \sim GP(0, \kappa)$
        \For{$t$ = 1, \dots, $T$}
            \State $x_t \leftarrow \argmin_{x \in \X} \mathrm{acq}_t(x)$
            \State Observe $y_t \gets f(x_t) +\varepsilon(x_t)$
            \State Update $GP$ posterior with $y_t$
        \EndFor
    \end{algorithmic}
\end{algorithm}
\vspace{-8 mm}

\subsection{Safe optimization under constraints}   

Consider a safety metric $q : \X \rightarrow \R$, which should stay bounded within a predefined region with high probability. Formally, the optimization problem is then \looseness-1
\begin{equation}
    \begin{aligned}
    \min_{x \in \X} \{ f(x)~|~ q(x) \leq c \},
\end{aligned}
\label{eq:constrained_bo}
\end{equation}
under the feasibility assumption $\{x \; | \; q(x) \leq c \} \neq \emptyset$, for a given $\textstyle{c \in \R}$.
The safety metric $q(x)$ here is an unknown, expensive-to-evaluate function that should be learned on the fly. \textsl{GoOSE} extends any standard BO algorithm providing high-probability safety guarantees in such a setup~\cite{koenig2021}. \looseness-1

\textsl{GoOSE} keeps track of two subsets of the optimization domain: \emph{the safe set} $\X_{t}$ and \emph{the optimistic safe set} $\X^{opt}_{t}$. $\X_{t}$ contains the set of points classified as safe while $\X^{opt}_{t}$ contains the points that potentially could be safe, i.e., 
\begin{align}
    \X_{t} &= \{x \in \X \; | \; \ucb^{q}_{t}(x) \leq c \},  \\
    \X^{opt}_{t} &= \{ x \in \X \; | \; \exists \bar{x} \in W_t, \text{ s.t. } g^{\epsilon}_{t}(\bar{x}, x) > 0  \}.
\end{align}
Here, the subscript $t$ indicates the iteration of the Bayesian optimization, and the superscript $q$ marks the constraint $q(x)$. $W_t \subset \X_{t}$ is the set of expanders - the periphery of the safe set such that $\textstyle{\forall \; x \in W_{t}}$ it holds true that $\textstyle{|\ucb^{q}_{t}(x) - \lcb^{q}_{t}(x)| > \epsilon}$, and $g^{\epsilon}_{t}(\bar{x}, x)$ is the noisy expansion operator taking values of $0$ or $1$, defined as: \looseness-1
\begin{equation}
g_{\epsilon}^{t}(\bar{x}, x)=\mathbb{I}\left[\lcb_{t}^{q}(\bar{x})+\left\|\mu_{t}^{\nabla}(\bar{x})\right\|_{\infty} d(\bar{x}, x)+\epsilon \leq c\right],
\label{eq:noisy_expansion_operator}
\end{equation}
for some $\textstyle{\epsilon > 0}$ uncertainty threshold relating to the observation uncertainty of $q(x)$ (see also \cite{koenig2021}). $\left\|\mu_{t}^{\nabla}(\bar{x})\right\|$ is the mean of the posterior over the gradient of the constraint function, and $d(\cdot, \cdot)$ is the distance metric associated with Lipschitz continuity of $q(x)$. Prior to the initialization of BO, \textsl{GoOSE} assumes a known non-empty safe seed $\X_{0}$, which does not need to be large in size and can be constructed from previous measurements. 

The acquired points belong to the joint domain $\X_{t} \cup \X^{opt}_{t}$ \looseness-1
\begin{equation}
    x_t = \argmin_{x \in \{ \X_{t} \cup \X^{opt}_{t} \}} \acq(x).
\end{equation}
If $x_t \in \X_{t}$, $f(x_t)$ and $q(x_t)$ are evaluated via available observations. Otherwise, if $x_t \in \X^{opt}_{t}$, then an expander of the safe set is evaluated. This strategy allows both to query the points satisfying the constraints and strategically expand the safe set in the direction of a promising input.

\subsection{Risk-averse optimization} 
\label{sec:risk-averse-opt}

Consider two different solutions that have similar expected function values but one produces noisier realizations (see \cref{fig:cost1d,fig:sigma1d} as a toy example). This is important when it comes to the actual deployment of the solutions in a real system where one might prefer more stable solutions. While standard risk-neutral BO methods, such as GP-LCB, provide guarantees in the risk-neutral (homoscedastic/heteroscedastic) BO setting \cite{Srinivas_2012}, they might fail in the risk-averse setting. To this end, \cite{makarova2021risk} proposes a practical risk-averse algorithm with theoretical guarantees. \looseness-1

\textbf{Mean-variance objective} is a simple and frequently used way to incorporate risk:
\looseness-1
\begin{equation}
    \min_{x \in \X}\{ MV(x) = f(x) + \alpha \rho^2(x) \}
    \label{eq:mean_var_obj}
\end{equation}
with $\alpha \geq 0$ is the \emph{coefficient of absolute risk tolerance}. Intuitively, the noise variance $\rho^2(x)$ acts as a penalizing term to the objective $f(x)$, encouraging to acquire points with less noise.  The coefficient of absolute risk-tolerance $\alpha$ allows accounting for various degrees of risk tolerance, with $\alpha=0$ corresponding to the risk-neutral case of using GP-LCB. \looseness-1

\textbf{Acquision function.} Since the noise variance is unknown in practice and both $f(x)$ and $\rho(x)$ are learned during the optimization, \textsl{RAHBO} proposes the acquisition strategy trading off not only exploration-exploitation but also risk: \looseness-1

\begin{equation}
    \mathrm{acq}_t(x) = \lcb^{f}_{t}(x) + \alpha \; \lcb^{var}_{t}(x),
    \label{eq:acq_f}
\end{equation}
where $\lcb^{var}_{t}$ denotes the lower confidence bound \cref{eq:lcb_ucb} constructed for the variance $\rho^2(x)$. For the latter, one need to observe evaluations $s^2_t$ of the noise variance, which is done by querying $k$ observations $\{y^{i}_{t}\}_{i = 1}^k$ of $f(x)$:\looseness-1
\begin{equation}
    \centering
    \begin{aligned}
    y_t =\frac{1}{k} \sum^{k}_{i=1} y^{i}_{t}, \quad 
    s^2_t = \frac{1}{k-1} \sum^{k}_{i=1}[y_t - y^{i}_t(x_t)]^2.
    \end{aligned}
    \label{eq:mean_var_measurements}
\end{equation}

For a complete introduction and treatment of \textsl{RAHBO}, we refer the reader to \cite{makarova2021risk}. Having introduced \textsl{RAHBO}, constituting an important part of the framework presented in this work, we are ready to present \textsl{RAGoOSE} -- our Safe Risk-averse Bayesian Optimization method.

\section{ \textsl{RAGoOSE}: Safe Risk-averse BO}\label{sec:method} 

\label{sec:ragoose}

We aim to minimize the risk-averse mean-variance objective subject to unknown safety constraint function. Similar techniques could be used to capture heteroscedastic noise in the constraint measurements.  Formally,
\begin{equation}
     \begin{aligned}
    \min_{x \in \X} \{ MV(x) ~|~ q(x) \leq c\},
\end{aligned}
\label{eq:Ragoose}
\end{equation}
for a given $c\in\R$. As before, we get noise-perturbed evaluations as follows: $m(x) = q(x) + \varepsilon_{q}$ with $\varepsilon_{q} \sim \mathcal{N}(0, \sigma_{q}^2)$, $\sigma_{q}^2 > 0,$ $y(x) = f(x) + \varepsilon(x)$ with $\varepsilon(x) \sim \mathcal{N}(0, \rho^2(x))$  and similarly $s^2(x) = \rho^2(x) + \varepsilon_{var}$ with $\varepsilon_{var} \sim \mathcal{N}(0, \sigma_{var}^2)$. The latter are observed  through sample mean $y_t$ and variance $s_t^2$ defined in \cref{eq:mean_var_measurements}. \looseness-1

\subsection{Building blocks of RAGoOSE}
 The main elements enabling the safe and risk-aware optimization of \textsl{RAGoOSE} rely on \textsl{RAHBO} and \textsl{GoOSE}. \looseness-1

\textbf{RAGoOSE probabilistic models.}
To reason under uncertainty about the unknown objective $f(x)$, noise variance $\rho^2(x)$ and safety metrics $q(x)$, we model each with a GP - $GP^{f}$, $GP^{var}$, and $GP^{q}$, respectively, characterised by a kernel function $\kappa^{f}, \kappa^{var}$, and $\kappa^{q}$, respectively. 

The posteriors of $GP_t^{var}$ and $GP_t^{q}$ can be simplified since the variance $\rho^2(x)$ and safety metrics $q(x)$ are assumed to be perturbed by homoscedastic noise:
\begin{align*}
    \mu^{q}_{t}(x) &= \kappa^q_t(x)^T(K^q_t + \sigma^2_{q} I)^{-1} y_{1:t}, \\
    {\sigma_{t}^{q}}^{2}(x) &= \kappa^q(x, x) - \kappa^q_t(x)^T(K^q_t + \sigma^2_{q} I)^{-1} \kappa^q_t(x), 
\end{align*}
where $I \in \R^{t \times t}$ is the identity matrix, and the rest corresponds to the notation in \cref{eq:GPmean,eq:GPvar}. The posterior mean and variance equations for $GP_t^{var}$ are built analogously.

The posterior $GP_t^{f}$ in \cref{eq:GPmean,eq:GPvar}, however, relies on the unknown noise variance $\rho^2(x)$ in $\Si_t$. To this end, \cite{makarova2021risk} proposes instead to use $\hat \Si_t$ with the upper confidence bound $\mathrm{ucb}^{var}_t(\cdot)$: \looseness-1
 \begin{align}\label{eq:hatsigma}
     \hat \Si_t := \tfrac{1}{k}\text{diag}\big(\mathrm{ucb}^{var}_t(x_1),\ldots, \mathrm{ucb}^{var}_t(x_t)\big),
 \end{align}
where $\hat \Si_t$ is corrected by $k$ due to the sample mean. This correction guarantees correct, though conservative, confidence bounds for the objective $f(x)$, thus allowing for effective optimization (see \cite{makarova2021risk} for more details). 

We assume the functions $f(x)$, $q(x)$, and $\rho^2(x)$ to belong to RKHS associated by their respective kernels and have bounded respective RKHS-norms, other assumptions follow \cite{makarova2021risk,koenig2021}. Under these assumptions, the lower and upper confidence bounds for the functions are defined according to \cref{eq:lcb_ucb} and denoted as  $\lcb^{q}_{t}(x), \ucb^{q}_{t}(x)$ for $q(x)$ and  $\lcb^{var}_{t}(x), \ucb^{var}_{t}(x)$ for $\rho^2(x)$. The corresponding parameters $\beta^f$, $\beta^q$, and $\beta^{var}$ are individually adjusted to ensure the validity of confidence bounds and balance exploration vs. exploitation (further referenced collectively as $\beta$ parameters). \looseness-1

\textbf{RAGoOSE acquisition strategy.}
With the confidence bounds defined, \textsl{RAGoOSE} acquires point $x_t$ at each iteration $t$ following the acquision strategy in \cref{eq:acq_f} conditioned on the safety of the inputs. Particularly,  $x_t$ is obtained by: \looseness-1
\vspace{-1 mm}
\begin{equation}
    x_{t} = \argmin_{x \in \{ \X_{t} \cup \X^{opt}_{t} \}} \{\lcb^{f}_{t}(x) + \alpha \; \lcb^{var}_{t}(x)\}
    \label{eq:opt_acq_ragoose}
\end{equation}

As described in \cite{makarova2021risk}, the framework allows for fine-tuning the trade-off between exploration, exploitation, and risk. This is exactly done via the parameter $\beta^{f}$, allowing to adjust exploration vs. exploitation trade-off of the objective, and $\alpha$, adjusting the trade-off between optimum value of the expected objective and the associated observation variance at this point. However, the parameters $\beta^{var}$ and $\beta^{q}$ act as tuning knobs to control exploration. As $\mathrm{acq}_{t}(x)$ is using $\lcb^{var}_{t}$, an increasing $\beta^{var}$ corresponds to a more optimistic belief on the observation noise variance of $f(x)$, whereas an increasing $\beta^{q}$ corresponds to a smaller safe set $\X_{t}$, and hence, to a more conservative estimate of the true safe set $\X_{safe}$ via $\X_{t}$, where $\X_{safe}$ is defined as $\X_{safe} = \{ x \in \X \; | \; q(x) \leq c \}$.
 \textsl{RAGoOSE} keeps track of its own estimate of the safe set $\X_{t}$ and the optimistic safe set $\X^{opt}_{t}$, defined in the same way as described in \autoref{sec:problem_statement}.

\subsection{The Algorithm}

\mypar{Initialization}
We assume the compact optimization domain $\X \subset \R^{d}$ to be defined and known, and an initial set of $n_{0}$ observations $\mathcal{D}_{f} = \{y_i\}_{i=1}^{n_0}$, $\mathcal{D}_{q} = \{m_i\}_{i=1}^{n_0}$, and $\mathcal{D}_{var} = \{s_i\}_{i=1}^{n_0}$ for $f(x)$, $q(x)$, and $\rho^2(x)$ at $x \in \X_{0} \subseteq \X_{safe} \subseteq \X$ must be available. We use $\X_{0}$ as a safe-seed set required for \textsl{RAGoOSE} initialization, and as a fallback input in case no safe inputs are found through \textsl{RAGoOSE} iterations. \looseness-1
Given the initialization sets, we initialize the three Gaussian processes $GP^{f}$, $GP^{var}$ and $GP^{q}$ with pre-selected kernels for the GPs  and tune their hyperparameters. \looseness-1

We then configure the \textsl{RAGoOSE} optimization with our selected values of $\alpha$ and $\beta$, defining the exploration-exploitation-risk aversion setting. Additionally, we select appropriate uncertainty threshold parameter $\epsilon$ governing the augmentation of the set of expanders, via the noisy expansion operator $g_{\epsilon}^{t}$. We also  set the number of evaluations per iteration $k$, and the total number of iteration loops of \textsl{RAGoOSE}, $T$. \looseness-1

\mypar{Optimization loop}
Having set the configuration and all parameters and input sets initialized, we present the main algorithm of our work, representing an end-to-end execution of \textsl{RAGoOSE}, in Algorithm \autoref{alg:ragoose_main}. We note that an explicit modeling of the variance in the constraint evaluations is not included, though a possible extension of the algorithm. Furthermore the expansion strategy (see Algorithm \autoref{alg:ragoose_main} line 12-15) is not risk-averse.

\begin{algorithm}
\begin{algorithmic}[1]
    \caption{\textsl{RAGoOSE} optimization loop}
    \State \textbf{Input}: Parameters $\alpha, \beta, k, \epsilon,$ 
    
    Initial data $ \X_{0}, \mathcal{D}_{f},\mathcal{D}_{var},\mathcal{D}_{q} $, 
    
    \text{Kernel functions} $\kappa^{f}, \kappa^{q}, \kappa^{var}$, 
    
    Prior means $\mu_0^f ,\mu^{q}_{0}, \mu_0^{{var}}$  \looseness-1

    \For{$t = 1, 2, \dots $} 
        \State update $GP^{f}$, $GP^{var}$, $GP^q$
        \State $\X_{t} \leftarrow\left\{x \in \mathcal{X}: u_{t}^{q}(x) \leq c\right\}$
        \State $L_{t} \leftarrow\left\{x \in \X_{t}: \exists z \notin \X_{t}\right.$, with $\left.d(x, z) \leq \Delta x\right\}$
        \State $W_{t} \leftarrow\left\{x \in L_{t}: u_{t}^{q}(x)-l_{t}^{q}(x) \geq \epsilon\right\}$
        \If{$x_{t-1} = x_{opt}$}
            \State $\textstyle{x_{opt} \leftarrow \argmin\limits_{\scriptstyle x \in \{\X_{t} \cup \X^{opt}_{t} \}} \{ \lcb^{f}_{t}(x) + \alpha\;\lcb^{var}_{t}(x)\}}$ 
        \EndIf
        \If{$\ucb^{q}_{t}(x) \leq c $} 
            \State $x_{t} \leftarrow x_{opt}$
            \State evaluate $y_t, m_t, s^2_t$
        \Else
            \State $x_{exp}$ $\leftarrow \argmin_{x \in W_{t}} d(x, x_{opt})$: $g^{\epsilon}_{t}(x, x_{opt}) \neq 0$ \looseness-1
            \State $x_{t} \leftarrow x_{exp}$
            \State evaluate $y_t, m_t, s^2_t$
        \EndIf
    \EndFor
    \State \textbf{Output}: $\textstyle{x^{*}_{final} \leftarrow \argmin\limits_{\scriptstyle x \in \X_{t} } \{\ucb^{f}_{t}(x) + \alpha \; \ucb^{var}_{t}(x)\}}$
    \label{alg:ragoose_main}
\end{algorithmic}
\end{algorithm}
\vspace{-5 mm}

\subsection{Implementation}
When optimizing the acquisition function, the current implementation of \textsl{RAGoOSE} uses particle swarm optimization (PSO) \cite{pso}, that initializes a set of particles across $\X_{t}$, and through its update rules also allows the particles to expand into the optimistic safe set $\X^{opt}_{t}$, avoiding calculating the optimistic safe set explicitly \cite{koenig2021}. \looseness-1

Another important aspect affecting the performance of the algorithm is the chosen discretization size $\Delta x$, as finer discretization allows a better assumption of $\X^{opt}_{t}$, due to a finer grid of expanders, while slowing down the computation time of the algorithm. We follow the  rule-of-thumb in selection of the discretization step from \cite{Duivenvoorden} based on the lengthscale parameter $\ell^q$ of the $GP^q$ kernel $\kappa^q$, with $\Delta x \text { s.t. } \kappa^{q}(x, x+\Delta x)\left(\sigma_{q}\right)^{-2}=0.95$.  \looseness-1

The parameter $\beta = 3$ was fixed for every GP of the implementation, resulting in a $99\%$ confidence interval. The exploration threshold was set to $\textstyle{\epsilon} = 6\sigma_q$ for all experiments (see also Algorithm \autoref{alg:ragoose_main} line 6).
\vspace{-0 mm}
\vspace{-3 mm}
\section{Case Study}
\label{sec:case_study}
We now apply the proposed approach, first on an illustrative example (sinusoidal function with varying noise levels), then on a numerical simulation of a precision motion system, and finally we run  experiments on the real system. In the synthetic example the hyperparemeters were tuned every iteration minimizing negative log marginal likelihood. The hyperparameters we used for the GPs in the other experiments can be read from:
\begin{table}[h]
    \centering
    \renewcommand{\arraystretch}{1.25}
    \resizebox{\columnwidth}{!}{
        \begin{tabular}{@{}c c c c@{}}
        \toprule
        Numeric experiment & $f$ & $q$ & $\rho^2$ \\ 
        \cmidrule(l{5mm}r{5mm}){1-4}
        $l = [\Pkp, \Vki]^T$ & $[50, 200]^T$ & $[50, 200]^T$ & $[50, 200]^T$\\
        $(\sigma_n,\mu_0,\lambda)$ & (2, 6, 0) & (4, 6, 1e-2) & (1e-2, 0.3, 1e-6 ) \\
        \midrule
        Real experiment & $f$ & $q$ & $\rho^2$ \\ 
        \cmidrule(l{5mm}r{5mm}){1-4}  
        $l = [\Pkp, \Vkp, \Vki, \Aff]^T$ & $[50, 100, 200, 0.5]^T$ & $[50, 100, 200, 0.5]^T$ & $[50, 100, 200, 0.5]^T$ \\
        $(\sigma_n,\mu_0,\lambda)$ & (1, 4.61, 0) & (1e-2, 0.192, 9e-4) & (1e-3, 1e-2, 1e-6 ) \\
        \bottomrule
        \end{tabular}%
        }
        \caption{GP Hyperparameters for experiments in \autoref{sec:case_study} 
        }
        \label{tab:gp_hyp}
\end{table}

All experiments use a RBF kernel $\kappa(x,x') = \sigma_n^2 exp(-\frac{(x - x')^2}{2l^2})$ for all GPs. All priors are set to a constant $\mu_0$. For the cost function $GP^f$ the noise is modeled by $GP^{var}$.

\begin{figure}[h]
     \centering
     \begin{subfigure}[b]{0.7\columnwidth}
         \centering
         \includegraphics[width=\textwidth]{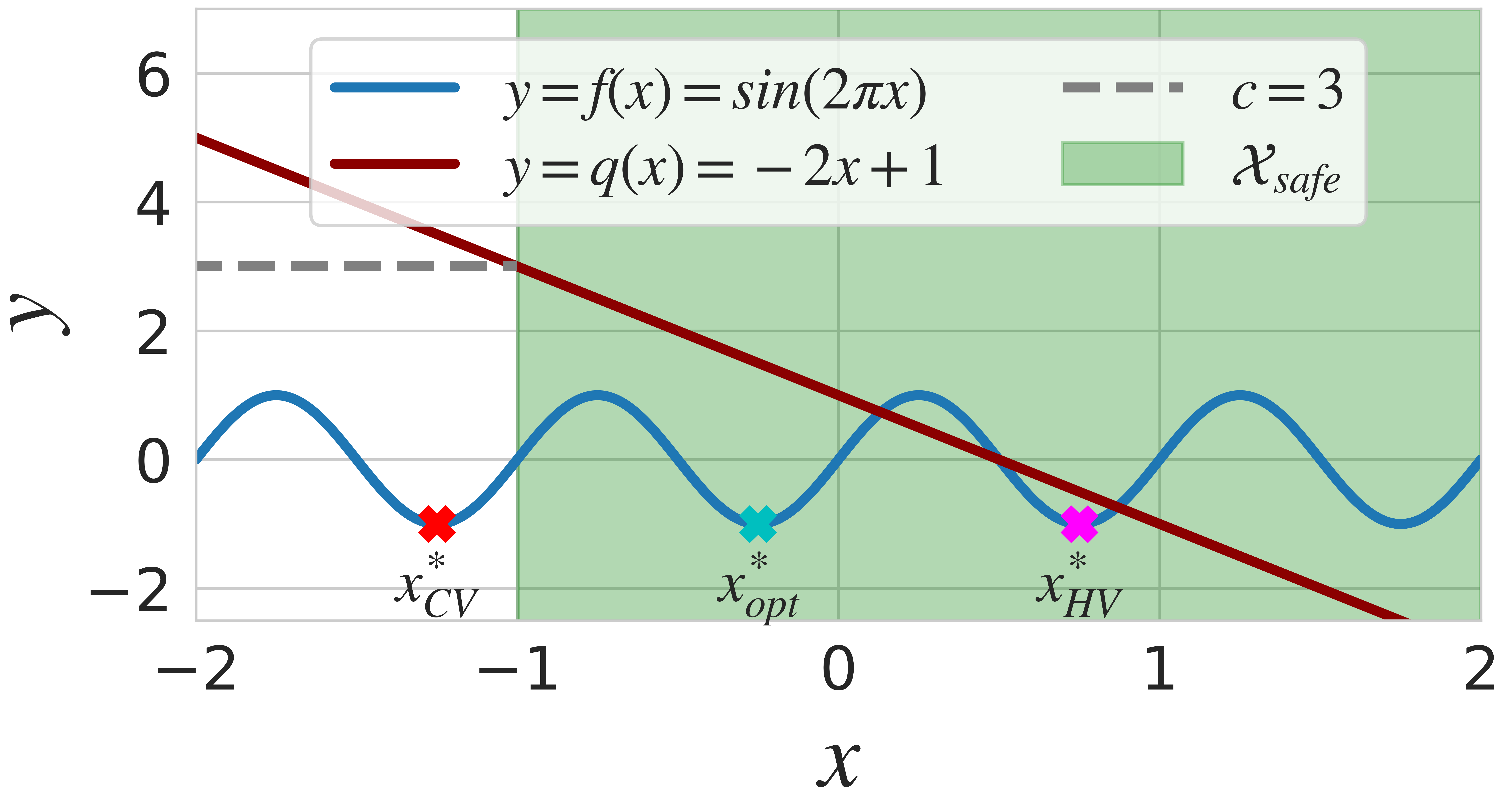}
         \label{fig:cost1d}
     \end{subfigure}
     \begin{subfigure}[b]{0.7\columnwidth}
         \centering
         \includegraphics[width=\textwidth]{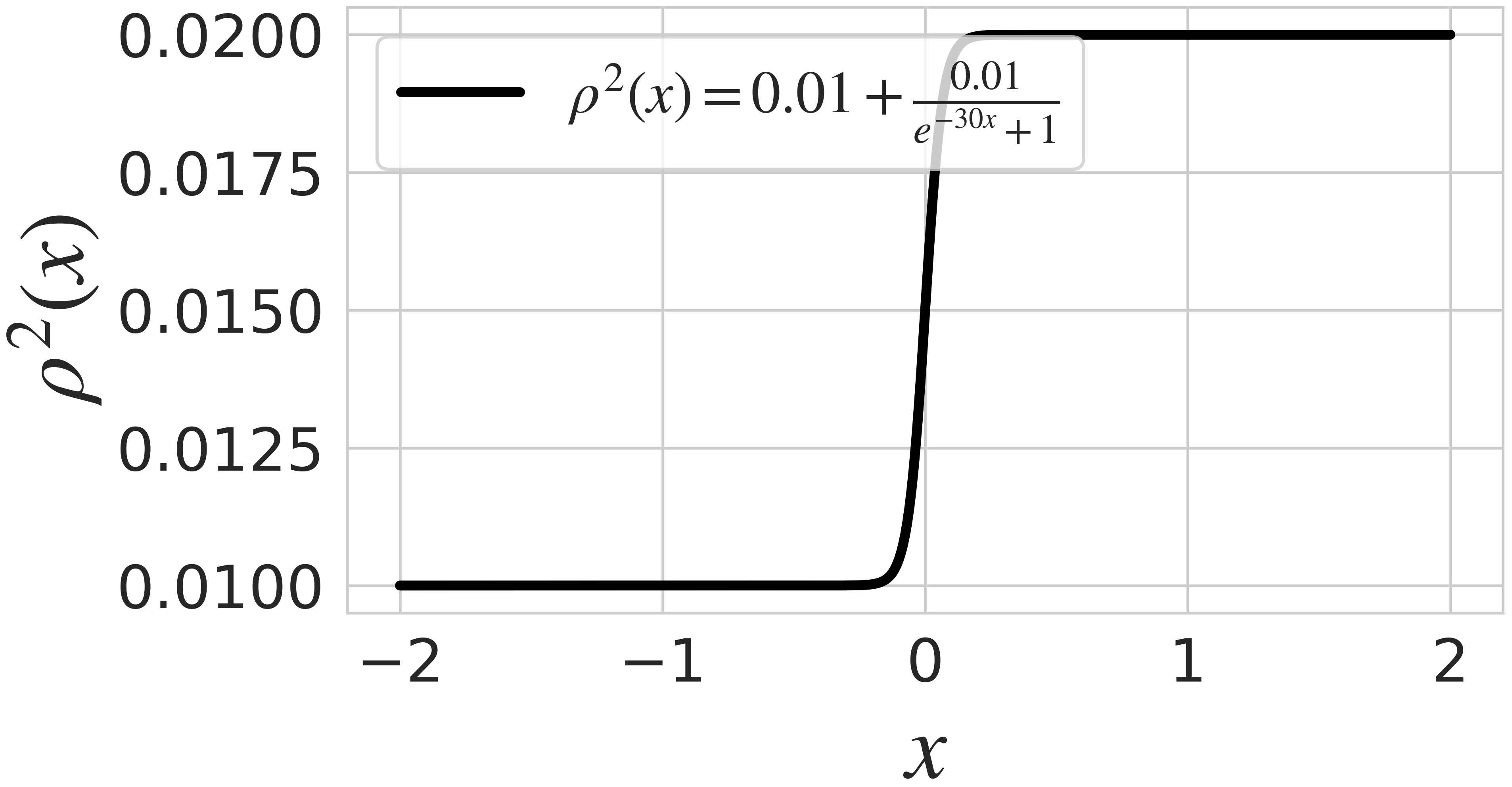}
         \label{fig:sigma1d}
     \end{subfigure}
              \vspace{-5 mm}

     \caption{Illustrative example in the case study. Top: Objective $f(x)$ with 3 global maxima marked as $(x_{CV}, x_{opt}, x_{HV})$ and constraint function $q(x)$ over the same domain showing that only two optima $(x_{opt}, x_{HV})$ are safe; Bottom: Heteroscedastic noise variance $\rho^2(x)$ over the same domain: the noise level at $(x_{CV}, x_{opt}, x_{HV})$ varies according to the sigmoid function.}
     \label{fig:1dsetup}
      \vspace{-0.1cm}
\end{figure}
      \vspace{-8 mm}

\subsection{Synthetic problem}
We study \textsl{RAGoOSE} on a synthetic optimization problem, and compare it with two baselines - \textsl{GoOSE} \cite{koenig2021} and Constrained Bayesian Optimization (\textsl{CBO}) \cite{Gardner}. \looseness-1

\begin{figure*}[hbt!]
    \centering
    \begin{subfigure}[b]{0.32\textwidth}
         \centering
         \includegraphics[width=\textwidth, trim=0cm 0cm 0cm 0cm ,clip]{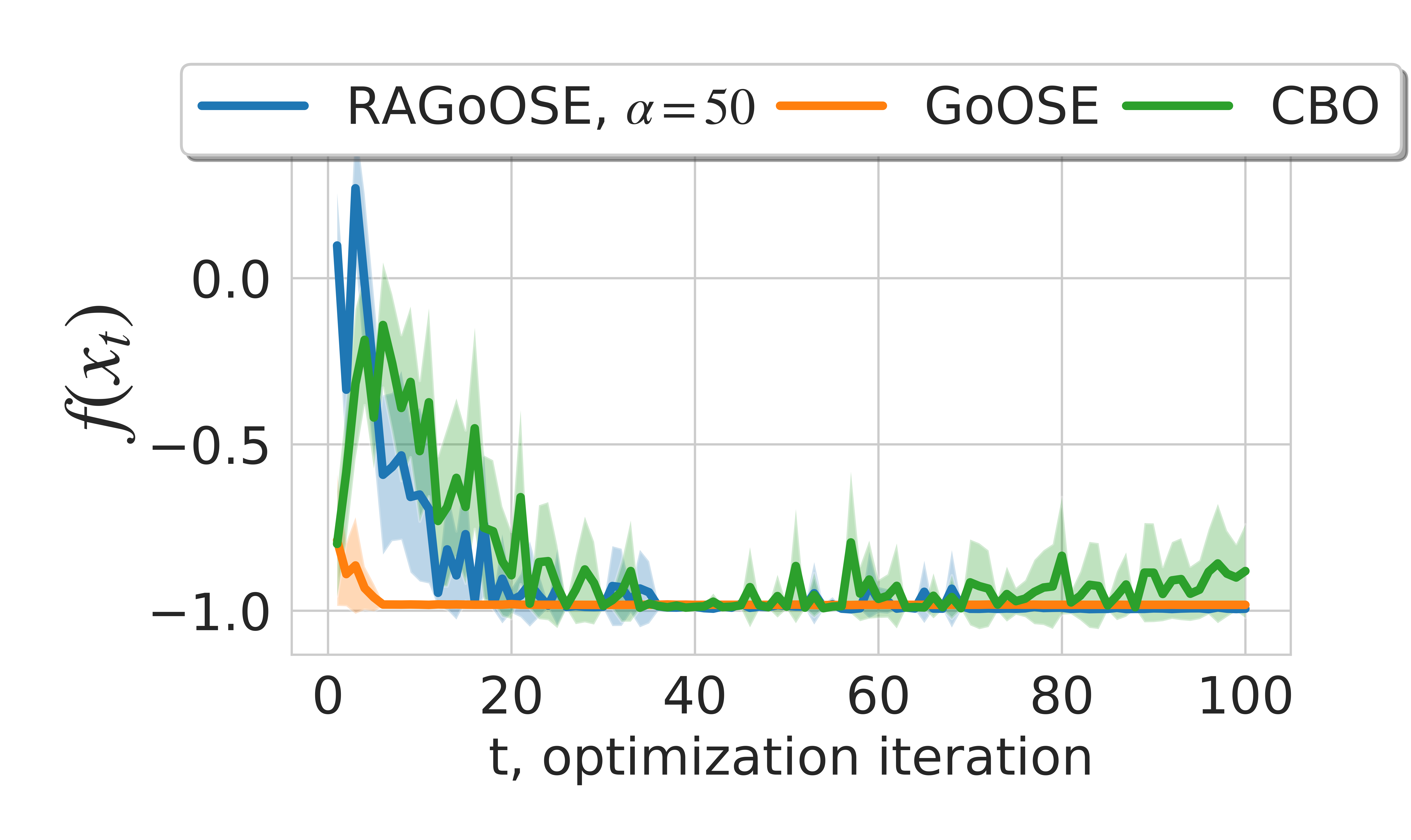}
         \vspace{-0.75cm}
         \caption{Cost function $f(x_{t}) \pm 2 SE$}
         \label{fig:1dcost}
    \end{subfigure}
    \begin{subfigure}[b]{0.32\textwidth}
         \centering
         \includegraphics[width=\textwidth]{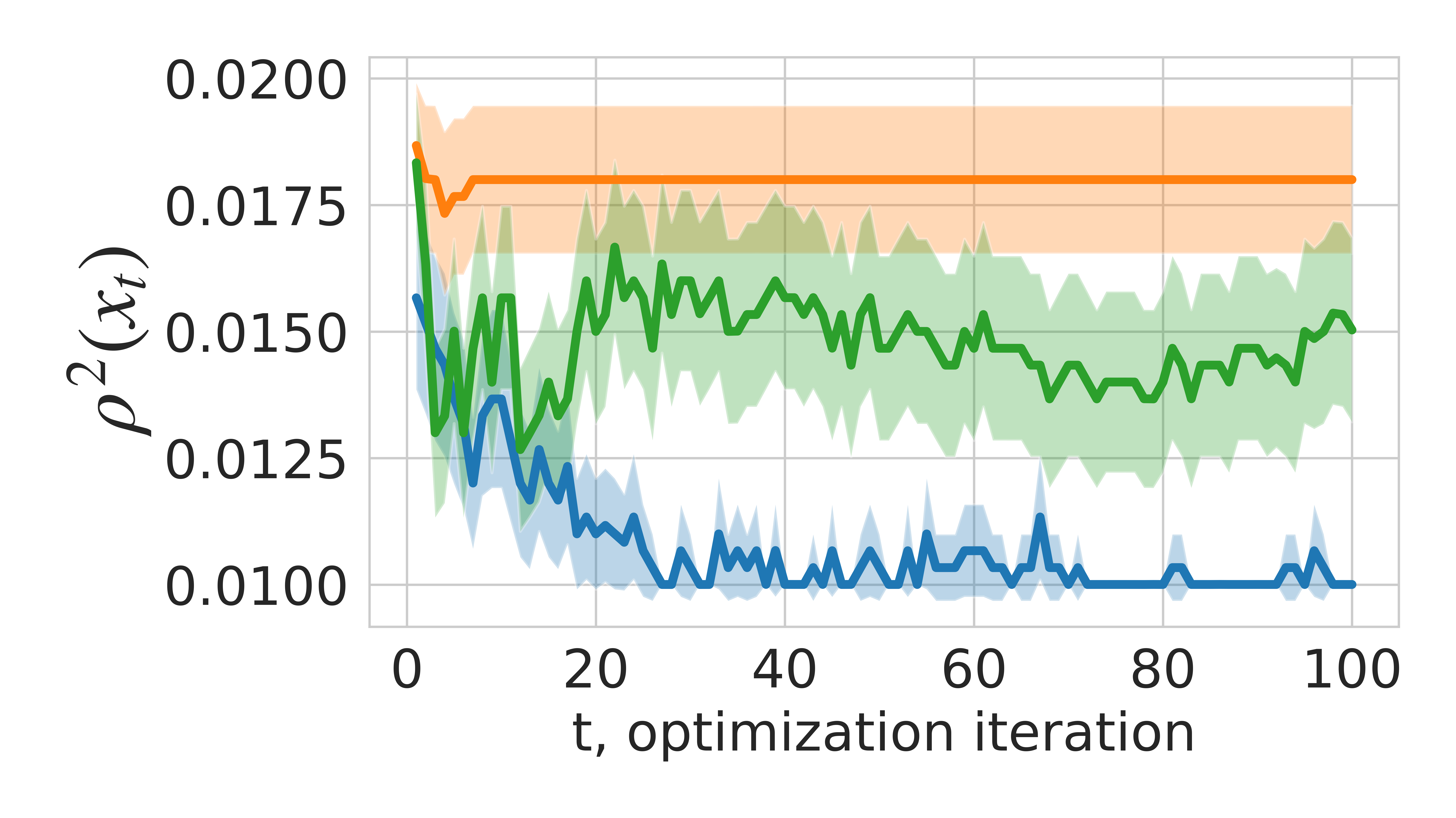}
         \vspace{-0.75cm}
         \caption{Noise variance $\rho^2(x_{t}) \pm 2 SE$}
         \label{fig:1dvar}
    \end{subfigure}
    \begin{subfigure}[b]{0.32\textwidth}
         \centering
         \includegraphics[width=\textwidth]{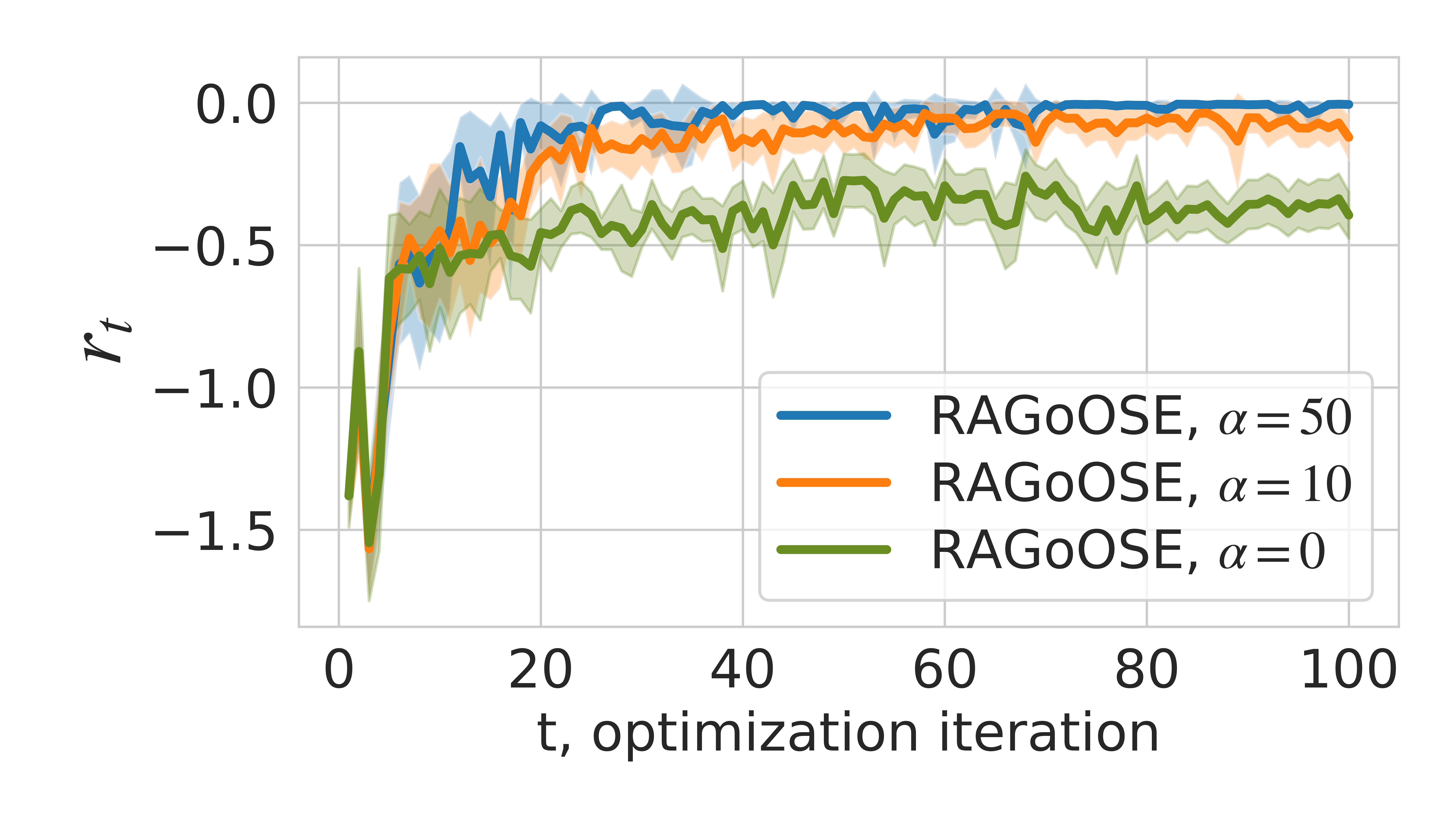}
         \vspace{-0.75cm}
         \caption{Regret $r_{t}$ }
         \label{fig:1dreg}
    \end{subfigure}
    \caption{Numerical study results for synthetic function (see \autoref{fig:1dsetup}), number of iterations $t=100$, and  $n_{rep}=30$ repetitions for each experiments, with reported average and two standard errors (SE). \textsl{RAGoOSE} successfully converges to the minimum of the cost function (a), while obtaining a lower observation noise at the solution (b) compared to the benchmarks. (c) Regret convergence as a function of the of $\alpha$ parameter.}
    \label{fig:1dexp}
    \vspace{-0.35cm}
\end{figure*}

    

\mypar{Problem Setup}
We use a sinusoidal cost function $f(x)$ and observe its noisy evaluations $y(x)$, with the cost observation noise variance function $\rho^2(x)$ as shown in \cref{fig:1dsetup}. Similarly, we observe the constraint function $q(x)$ via its noisy observations $m(x)$, with a homoscedastic noise variance of $\sigma_q=0.1$. Performance was assessed by repeating the optimization 30 times, with $T=200$ BO iterations each, and $k=10$ evaluations of the cost function. We repeat the optimization routine with \textsl{RAGoOSE} for all $\alpha \in \{0, 10, 30\}$. As \cref{fig:cost1d} illustrates, the set-up gives rise to 3 optimizers - $x^{*}_{CV}$, that violates the constraint, $x^{*}_{opt}$, located at the low observation noise variance region of the domain, and $x^{*}_{HV}$, located in the high-variance region. All GPs were initialized with five a priori known points within the safe set interval $x \in [0, 1]$. In \cref{fig:1dexp} we report the mean cost function $f(x_t)$, observation noise variance $\rho^2(x_t)$ with the best performing \textsl{RAGoOSE} configuration and the benchmarks, and the intermediate regret $r_{t}$ with varying $\alpha$ (with $r_{t}$ defined in \cref{eq:regret}) with $\pm 2$ standard errors ($SE$) across the $n=30$ performed repetitions of the optimization routine
\begin{equation}
    r_{t} = \left[f(x^{*}_{opt}) + 50\rho^2(x^{*}_{opt})\right] - \left[f(x_t) + 50
\rho^2(x_t)\right] .
    \label{eq:regret}
\end{equation}
We report a constraint violation when the observed value $m_t$ exceeds the constraint upper bound $c=3$. \looseness-1

\mypar{Results} All methods were able to converge to the optimum of the cost function successfully, as shown in \cref{fig:1dexp}. While  \textsl{RAGoOSE} requires more iterations to converge, it shows superior performance in observation noise variance, converging to $x^{*}_{opt}$, yielding a 41\% and 31\% decrease in observed noise variance, compared with \textsl{GoOSE} and \textsl{CBO} respectively. We observe a closer convergence to the low-variance optimum with increasing $\alpha$, as \cref{fig:1dreg} reports, with $\alpha=50$ yielding the lowest regret at the terminal iteration. As expected, the risk-neutral case of $\alpha=0$ fails to converge to the risk-averse optimum $x^{*}_{opt}$, yielding the highest intermediate regret $r_{t}$. \textsl{RAGoOSE} outperforms the other benchmarks in the number of constraint violations. The benchmark study results are summarized in \cref{tab:1dres}. The additional computation due to modeling the noise variance slows down the optimization time of \textsl{RAGoOSE}, but increases the performance and reduces constraint violations. Furthermore \textsl{RAGoOSE} shows slightly better prediction of the final optimum mean value, within one standard deviation of the noise.

\begin{table}[h]
\centering
    \renewcommand{\arraystretch}{1.25}      
        \resizebox{0.95\columnwidth}{!}{%
        \begin{tabular}{@{}c c c c@{}}
        \toprule
                  \textit{Metric/Parameter}                 & \textit{RAGoOSE}  & \textit{GoOSE} & \textit{CBO}  \\
                                      	\cmidrule(lr){1-4}
        $\overline{f(x^{*})}$      & \textbf{-0.999} & -0.987       & \textbf{-0.999}      \\ 
        $\overline{\rho^2(x^{*})}$ & \textbf{0.011}   & 0.018         & 0.016        \\ 
        Constraint violations         & \textbf{0.03\%}   & 0.07\%         & 7.90\%        \\ 
        Mean optimization time [s]         & 68.6              & 28.4           & \textbf{21.4} \\
        \bottomrule
        \end{tabular}%
        }
        \caption{Mean optimized performance metric (cost) $\overline{f(x^{*})}$, mean noise parameter $\overline{\rho^2(x^{*})}$, constraint violations and optimization time for different BO-based tuning methods applied on the synthetic problem across the 30 optimization experiments}
        \label{tab:1dres}
\end{table}

        \vspace{-3 mm}
\subsection{Controller Tuning for Precision Motion}

\begin{figure}
\centering
  \begin{subfigure}[h]{0.7\columnwidth}
    \centering
    \includegraphics[width=\linewidth]{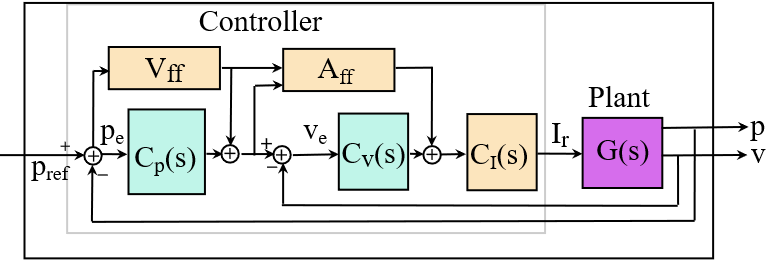}
    \label{fig:Scheme} 
    \end{subfigure}
  \begin{subfigure}[h]{0.7\columnwidth}
  \centering
    \includegraphics[width= 0.9\linewidth]{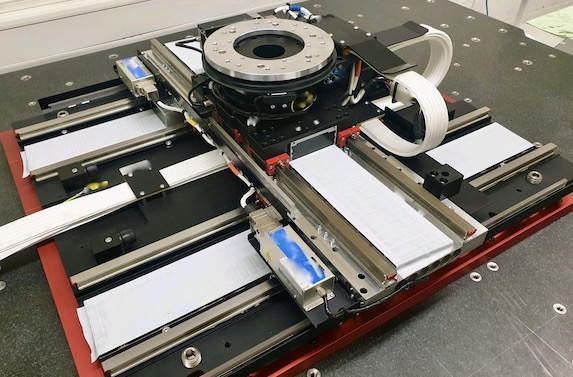}
  \end{subfigure}
  \caption{Top panel: Simplified controller architecture in the experimental study. Bottom panel: Positioning system by Schneeberger Linear Technology AG. }
  \label{fig:argussetting}
\end{figure}

\mypar{System and controller}
The system of interest is a linear axis of a high precision positioning system by Schneeberger Linear Technology, shown in \cref{fig:argussetting}. The axis is driven by a permanent magnet AC motor equipped with nanometer precision encoders for position and speed tracking. The axis is guided on two rails, reaching positioning accuracy of $\leq 10\,\mathrm{\mu m}$, repeatability of $\leq 0.7 \,\mathrm{\mu m}$, and $3\sigma$ stability of $\leq 1 \,\mathrm{nm}$.
The sampling time of the controller and the date acquisition of the system used is $f_s = 20\,\mathrm{kHz}$.
Such systems are routinely used for production and quality control in the semiconductor industry, in biomedical engineering, and in photonics and solar technologies.
For example, in semiconductor manufacturing applications, considering risk or safety alone is not enough due to stringent production requirements.
Therefore a tuning method that considers both process variance and safety, e.g., \textsl{RAGoOSE}, is beneficial.

%
%


The system is controlled by a three-level cascade controller shown in \cref{fig:Scheme}.
The outermost loop controls the position with a P-controller $\textstyle{\Cp(s)=\Pkp}$, and the middle loop controls the velocity with a PI-controller $\textstyle{\Cv(s) = \Vkp + \Vki/s}$. 
The inner loop, which controls the current of the drive, is well-tuned and not subject to tuning. Feedforward structures are used to accelerate the response of the system.
The gain of the velocity feedforward $\Vff$ is well-tuned and not modified during the retuning procedure, while the acceleration feedforward gain $\Aff$ is retuned by the algorithm during the experiments on the real system and fixed to $\Aff = 0$ for the simulation experiments. 

\mypar{Experimental setup}
We optimize the controller parameters providing the best tracking performance of the system, while avoiding instabilities of the controller in the presence of heteroscedastic noise. We estimate both the tracking performance and potential instabilities via the corresponding data-driven features $\phi(x)$ and $q(x)$ extracted from the system signals. The optimization problem can be written as follows:
\begin{align}
    \min_{\mathrm{\bm{x}}\in\mathcal{X}}\{\phi(\mathrm{\bm{x}})~|~q(\mathrm{\bm{x}})>c \}. \nonumber
\end{align}
Specifically, the performance tracking metric is provided by $\phi(\mathrm{\bm{x}})= \frac{1}{n_{P} - n_{s}} \textstyle{\sum_{i=n_{s}}^{n_{P}} |\xi(i, n_{s})\mathrm{p_e}(t_{i})|}+\delta(\mathrm{\bm{x}})$. Here, $\mathrm{p_e}(t_i)$ indicates the deviation from the reference position at each time instance $t_i$, $\delta$ indicates if additional noise is added to the metric (only used in the numerical study). The stability (safety) constraint $q(\mathrm{\bm{x}})$ is calculated from the Fourier transform of the filtered velocity error $\mathcal{F}[\mathrm{v_e}(t_{i})]$ and limited by a threshold $c$. The threshold is determined by a small number of  prior experiments. Furthermore, in the experimental study, both the performance and the stability metrics are filtered by a right-sided sigmoid function $\xi(i,n_i)$ centered at $150\text{ms}$ after the start of the settling phase for robustness of the metric. The settling phase extends from time instance $n_s$ to time instance $n_p$. The optimization is over the controller gains $\Pkp$, $\Vkp$, $\Vki$ and the feedforward gain $\Aff$.

\begin{figure}
\centering
  \begin{subfigure}[h]{0.48\columnwidth}
    \centering
    \includegraphics[width=\linewidth]{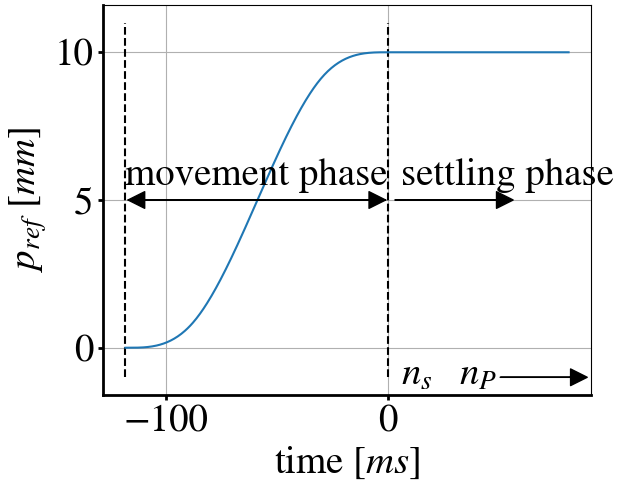}
    \label{fig:pref} 
    \end{subfigure}
  \hfill
  \begin{subfigure}[h]{0.48\columnwidth}
  \centering
    \includegraphics[width=\linewidth]{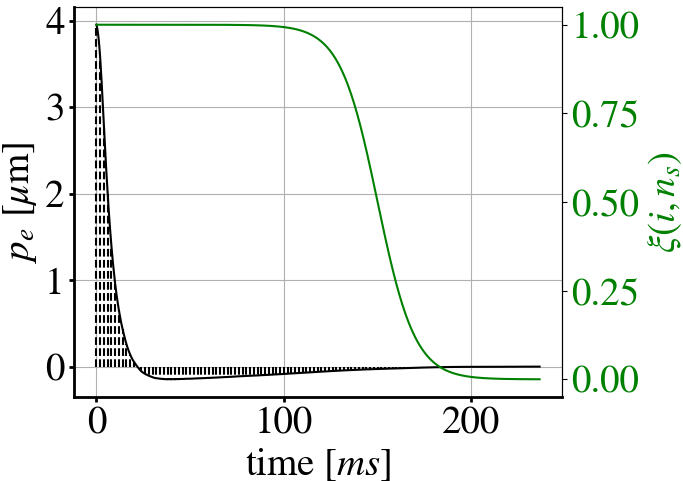}
    \label{fig:pe}
  \end{subfigure}
  \vspace{-5 mm}
  \caption{Left panel: S-curve position reference function indicating move- and settle time. Right panel: position error $p_e(t_i)$ example (highlighted in black) and $\xi(i, n_s)$ filter}
  \label{fig:cost_function}
\end{figure}

We solve the optimization problem using \textsl{RAGoOSE}, following \eqref{eq:Ragoose} and Algorithm \ref{alg:ragoose_main}.
As our system exhibits heteroscedastic noise which is included in $\phi(\mathrm{\bm{x}})$, to provide the model for its variance i.e. $\rho^2(\mathrm{\bm{x}})$, we need to acquire multiple repetitions at each candidate gain combination $\mathrm{x}$.
For observations of $f(\mathrm{x})$, we use the mean of the observations over the $n$ repetitions, while we use the variance for $\rho^2(\mathrm{x})$. 
The optimum of the \textsl{RAGoOSE} experiments is found using the pessimistic prediction given by:\looseness=-1
\begin{equation}
    x^* = \argmin_{\mathrm{\bm{x}} \in \X_{t}} \{\ucb^{f}_{t}(\mathrm{\bm{x}}) + \alpha \; \ucb^{var}_{t}(\mathrm{x})\} .
    \label{eq:pessimistic_prediction}
\end{equation} 

\begin{figure}[H]
    \centering
    \begin{subfigure}[b]{0.55\columnwidth}
         \centering         \includegraphics[width=\linewidth]{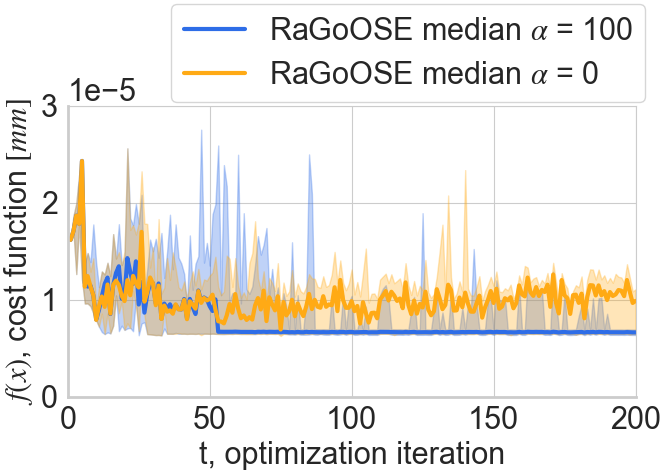}
         \label{fig:argus_sim_ragoose_mean}
    \end{subfigure}
    \vspace{-2 mm}
    \begin{subfigure}[h]{0.55\columnwidth}
         \centering         \includegraphics[width=\linewidth]{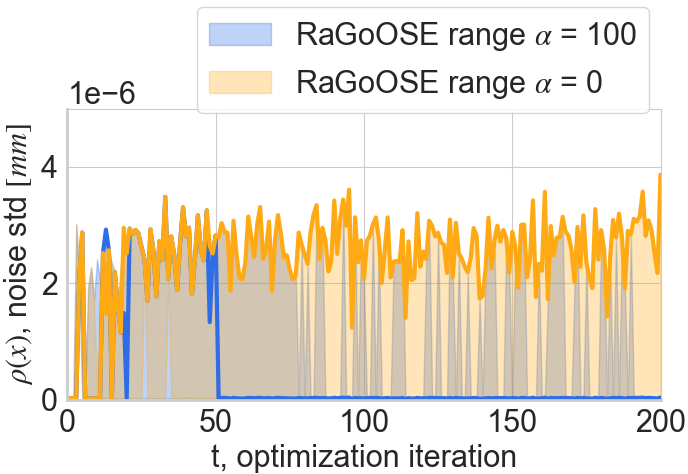}
        \vspace{-0.5 cm}
         \label{fig:argus_sim_ragoose_std}
    \end{subfigure}
    \caption{Optimization results (numerical) for controller tuning comparing \textsl{RAGoOSE} using $\alpha=0$ with $\alpha=100$ for 10 different experiments. (a) Median and min-max interval of $f(\mathrm{x})$ over the 10 \textsl{RAGoOSE} optimizations for each iteration. (b) Median standard deviation of the cost $\rho(x)$ over the 10 repeated \textsl{RAGoOSE} optimizations for each iteration together with its min-max interval.}
    \label{fig:argussim}
\end{figure}

Each experiment has ten repetitions for each candidate combination of parameters $\mathrm{x}$, to estimate the noise variance $\rho(\mathrm{x})$. Furthermore, we repeat the \textsl{RAGoOSE}  optimization ten times. Figure \ref{fig:argussim} shows the range of values for $f(\mathrm{x})$ and $\rho(\mathrm{x})$, obtained for each iteration during the ten repeated optimizations. The solid lines show the median over the repeated optimizations.
\vspace{-4 mm}

\mypar{Numerical study}
We perform the \textsl{RAGoOSE} optimization on a simulation of a single axis of Argus comparing the effect of $\alpha=0$ and $\alpha=100$ with $T=200$ Bayesian optimization iterations and $n=10$ evaluations of the cost function at each iteration. For simplicity we only optimize the gains $x=[\Pkp, \Vki]$ in the simulation experiment. All \textsl{RAGoOSE} experiments are initialized using 3 initial samples, $x_0 = [[200, 1000], [300, 1000], [200, 1500]]^T$. To simulate the heteroscedasticity of the observations, non uniform noise is added on the cost function, dependent on one of the two parameters, e.g. for $\mathrm{x_1}=\Vki$,
\begin{equation}
\vspace{-2mm}
\delta(\mathrm{x_1}) = \begin{cases}
  \mathcal{N}(0,1e-7) , \mathrm{x_1} \leq 1200 \\
    \mathcal{N}(0,1e-5)  , \mathrm{x_1} > 1200 .
    \end{cases}
\end{equation}

Each experiment has ten repetitions for each candidate combination of parameters $\mathrm{x}$, to estimate the noise variance $\rho(\mathrm{x})$. Furthermore, we repeat the \textsl{RAGoOSE}  optimization ten times. Figure \ref{fig:argussim} shows the range of values for $f(\mathrm{x})$ and $\rho(\mathrm{x})$, obtained for each iteration during the ten repeated optimizations. The solid lines show the median over the repeated optimizations.
During the simulation experiment of the Argus linear axis, \textsl{RAGoOSE} using $\alpha=0$ and $\alpha=100$ show significant differences in performance and observation noise (see \autoref{fig:argussim}). While both \textsl{RAGoOSE} configurations manage to quickly find a parameter combination resulting in good performance, but high risk, \textsl{RAGoOSE} using $\alpha=100$ manages to further explore and find a combination resulting in lower risk and thus also better performance, due to the effect of the noise on the cost observation.

\begin{figure}[h]
    \centering
    \begin{subfigure}[h]{0.47\columnwidth}
         \centering
         \includegraphics[width=\linewidth]{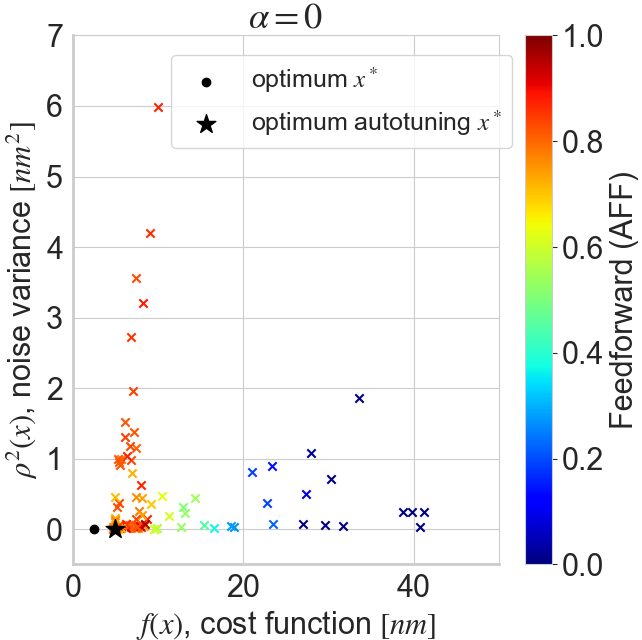}
    \end{subfigure}
    \vspace{-0 mm}
    \begin{subfigure}[h]{0.47\columnwidth}
         \centering
         \includegraphics[width=\linewidth]{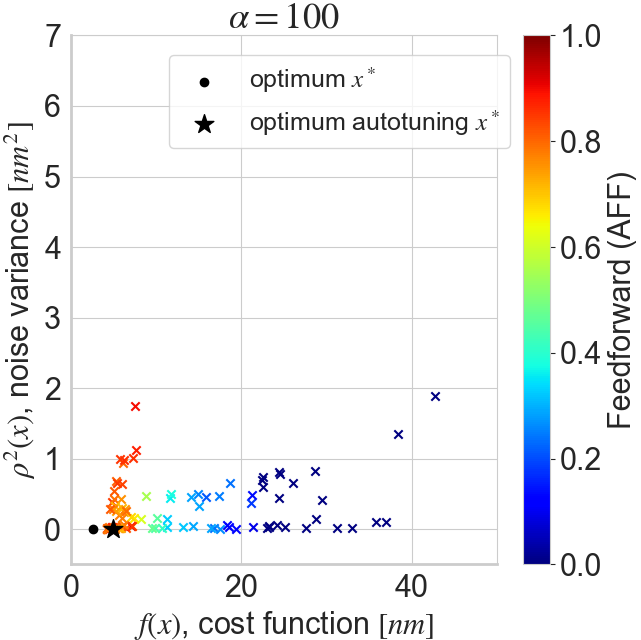}
        \vspace{-2 mm}
    \end{subfigure}   
    \caption{Optimization results for controller tuning using \textsl{RAGoOSE} with, $\alpha=0$ (left) and  $\alpha=100$ (right). Mean observation over the 10 repetitions $f(\mathrm{x})$ (x-axis) is plotted against the noise observation $\rho^2(x)$ (y-axis), for one \textsl{RAGoOSE} optimization. The resulting data points are color-coded corresponding to the values of one of the optimization variables, $A_{\mathrm{ff}}$, to visualize the heteroscedasticity of the noise, which induces the particular shape of the obtained distribution with distinct "branches", corresponding to high or low noise.}
    \label{fig:argusexp}
\end{figure}

\mypar{Experimental Study}
We perform the optimization with \textsl{RAGoOSE} using $\alpha=0$ and $\alpha=100$ with $T=100$ Bayesian optimization iterations and $n=10$ evaluations of the cost function at each iteration. We show the distribution of the observations of the mean $f(x_t)$ and the observations of the variance $\rho^2(x_t)$ of $\phi(x)$ in \autoref{fig:argusexp}.
\vspace{-0 mm}

We further compare the performance of \textsl{RAGoOSE} using $\alpha=0$ and $\alpha=100$ for controller tuning of the three  controller gains ($\Pkp$, $\Vkp$, $\Vki$) and on the acceleration feedforward parameter ($\Aff$) with the performance of the auto-tuning routine of the built-in hardware controller balancing bandwidth with phase and gain margin method on the real Argus linear axis. Both \textsl{RAGoOSE} optimizations were stopped after 100 iterations. The first point of each of the two data-driven optimizations corresponds to the controller parameters found by manual tuning, with $x_0 = \left[200, 600, 1000, 0.0\right]^\top$. Figure \ref{fig:argusexp} shows the  distribution of cost mean observations $f(x_t)$ with the corresponding variance observation $\rho^2(x_t)$. Clearly,  $\alpha=100$ avoids high risk parameter combinations compared to using $\alpha=0$ and the preferred candidate controllers are along low variance regions, corresponding to lower values of the $A_{ff}$ parameter which is mostly associated with heteroscedastic noise.
\vspace{-0 mm}
 \begin{table}[h]
    \centering
    \renewcommand{\arraystretch}{1.25}
    \resizebox{0.95\columnwidth}{!}{%
        \begin{tabular}{@{}c c c c@{}}
        \toprule
        & RaGoOSE \textit{$\alpha=0$} & RaGoOSE \textit{$\alpha=100$} & Auto-tuning \\ 
        \cmidrule(lr){1-4}
        $f(x^*) [nm]$ & 2.459 & 2.543 & 4.987\\ 
        $\rho^2(x^*) [nm^2]$ & 9.187e-3 & 8.347e-3 & 5.832e-3\\
        $\Pkp^*$ & 330.4 & 319.8 & 350.0\\
        $\Vkp^*$ & 731.6 & 862.4 & 600.0\\
        $\Vki^*$ & 1449.3 & 1174.6 & 2000.0\\
        $\Aff^*$ & 0.83 & 0.815 & 0.0\\
        \bottomrule
        \end{tabular}%
        }
        \caption{Optimized performance metric (cost), noise parameter, and corresponding controller parameters for different values of $\alpha$, compared to the controller's inbuilt auto-tuning in the experimental study.}
        \label{tab:argusres}
\end{table}
\vspace{-2 mm}
\looseness=-1
The performance of the controllers corresponding to optima found by \textsl{RAGoOSE} using $\alpha=0$ and $\alpha=100$ (see \autoref{fig:argusexp}) after $t=100$ iterations is similar in tracking performance (cost) and in variance. This is a result of the specific heteroscedasticity of the system's noise. Due to the distribution of the noise, the parameter regions of optimal performance correspond to the parameter regions of low risk, situated in the bottom left corner of the graphs in Figure \ref{fig:argusexp}. Thus, the final result of the RaGoOSE optimization is not strongly dependent of the choice of $\alpha$. Both methods show significant improvement compared to the controllers tuned using the built-in auto-tuning routine (see \autoref{tab:argusres}), due to the contribution of the noise model in the performance modelling. There were no constraint violations observed during the optimization of both \textsl{RAGoOSE} experiments. In terms of duration of the tuning, \textsl{RAGoOSE} needs ten times more evaluations, to estimate the noise variance, which slows it with one order of magnitude, compared to methods not considering heteroscedastic noise. In addition, as shown in Table \ref{tab:1dres}, it is slower due to the additional GP model of the noise. Thus, one iteration of \textsl{RAGoOSE} for tuning the precision motion system takes $32\mathrm{s}$, compared to $1.6\mathrm{s}$ achieved with the method from \cite{koenig2021}.
 \vspace{-1mm}



\vspace{-1 mm}
\section{Concluding remarks}
\looseness=-1
We presented a risk-averse Bayesian optimization framework that performs constrained optimization with safe evaluations, accounting for the solution variance. To this end, we considered heteroscedastic noise in the process and extended the RAHBO method to consider the safety of evaluations in terms of constraint violations. We present results in both numerical benchmark studies and on a real high-precision motion system to show the improved performance over state-of-the-art methods.  We demonstrate the proposed framework on a controller tuning application where safe evaluations are crucial to ensure physical safety. Future work will consider extending \textsl{RAGoOSE} to include the variance from the measurements of the constraints in the optimization. From the application perspective, the next challenge is applying the method in continuous optimization scenarios, without disrupting the operation of the physical system.
\vspace{-4 mm}

\bibliography{bib,report_bibs}
\bibliographystyle{IEEEtran}


\end{document}